\documentclass[apj]{emulateapj}
\usepackage{graphicx}
\usepackage{amsmath,natbib}
\usepackage{ulem}

\usepackage{color}

\newcommand{\feoh}{[{\rm Fe} / {\rm H}]}
\newcommand{\msun}{\, M_\odot}
\newcommand{\Zsun}{\, Z_\odot}
\newcommand{\Tv}{T_{\rm vir}}
\newcommand{\mmd}{M_{\rm md}}

\def\mhyph{\, \mathchar`- \, }
\def\abra#1#2{[{\rm #1}/{\rm #2}]}
\def\mmdp#1{M_{\rm md, #1}}

\shorttitle{Pop~III stars}
\shortauthors{Komiya et al.}

\begin{document}
\title{Population III stars around the Milky Way}

\author{Yutaka Komiya\altaffilmark{1}, Takuma Suda\altaffilmark{1} and Masayuki Y. Fujimoto\altaffilmark{2}\altaffilmark{3}}
\altaffiltext{1}{Research Center for the Early Universe, University of Tokyo, Hongo 7-3-1
Bunkyo-ku, Tokyo 113-0033, Japan}
\altaffiltext{2}{Department of Cosmoscience, Hokkaido University, Sapporo, Hokkaido 060-0810, Japan}
\altaffiltext{3}{Department of Engineering, Hokkai-Gakuen University, Sapporo, Hokkaido 062-8605, Japan}

\begin{abstract}
We explore the possibility of observing Population III (Pop~III) stars, born of the primordial gas. 
   Pop~III stars with masses below $0.8 \msun$ should survive to date though are not observed yet, but the existence of stars with low metallicity as $\feoh < -5$ in the Milky Way halo suggests the surface pollution of Pop~III stars with accreted metals from the interstellar gas after birth.  
In this paper, we investigate the runaway of Pop~III stars from their host mini-halos, considering the ejection of secondary members from binary systems when their massive primaries explode as supernovae.  
   These stars save them from the surface pollution.  
   By computing the star formation and chemical evolution along with the hierarchical structure formation based on the extended Press--Schechter merger trees, we demonstrate that several hundreds to tens of thousands of low-mass Pop~III stars escape from the building blocks of the Milky Way.  
   The second and later generations of extremely metal-poor (EMP) stars are also escaped from the mini-halos.  
   We discuss the spatial distributions of these escaped stars by evaluating the distances between the mini-halos in the branches of merger trees under the spherical collapse model of dark matter halos. 
   It is demonstrated that the escaped stars distribute beyond the stellar halo with a density profile close to the dark matter halo, while the Pop~III stars are slightly more centrally concentrated . 
 Some escaped stars leave the Milky Way and spread into the intergalactic space.  
 Based on the results, we discuss the feasibility of observing the Pop~III stars with the pristine surface abundance. 
\end{abstract}

\section{Introduction}

The first generation stars in the context of the cosmic chemical evolution are expected to be metal-free, and are referred to as Population~III (Pop~III) stars. 
One of the key questions in the galactic archeology is whether or not there are low-mass Pop~III stars. 
If some of Pop~III stars are born with low mass that have survived until now, they may still shine as nuclear burning stars in the solar vicinity. 
We refer to Pop~III stars with such low masses ($< 0.8 \msun$) as Pop~III survivors in this paper.

In the context of star formation, the mass range of Pop~III stars has been investigated by many computational studies. 
Recent numerical simulations of the first star formation have proposed pathways to form low-mass Pop~III stars.  
These studies reveal that 
a primordial gas cloud fragments to form a binary \citep[e.g.][]{Machida08, Turk09, Stacy10, Stacy13} or a star cluster \citep[e.g.][]{Clark11, Greif11, Susa14} without metals.  
The predicted mass functions of the Pop~III protostars extend down to $\sim 0.8 \msun$ or below \citep{Clark11, Greif11, Stacy13}.  
   Some protostars may remain low-mass stars depending on the efficiency of later gas accretion and merging of protostars after the end of the computations 
\citep{Greif11, Smith11}.

Observationally, 
on the other hand, 
 no stars with zero surface metallicity have ever been found yet despite intensive
large-scale surveys in the Milky Way (MW) halo (e.g., HK survey, Beers et al.\ 1992; Hamburg/ESO[HES] survey, Christlieb et al.\ 2001; SDSS/SEGUE, Yanny et al.\ 2009; and SkyMapper Southern Sky Survey, Keller et al.\ 2007) 
These surveys have found thousands of candidates of extremely metal-poor (EMP) stars. 
Several hundreds of stars with $\feoh < -3$ have received follow up observations with high-resolution spectroscopy. 
Among these EMP stars, there are five hyper metal-poor (HMP) stars with $\feoh \leq -5$ and four additional ultra metal-poor (UMP) stars with $-5 <  \feoh < -4.5$ but no star with $Z=0$.   
The most iron-deficient star \citep[SMSS J031300.36-670839.3,][hereafter SMSS J0313-6708]{Keller14} shows no evidence of iron \citep[$\feoh < -7.52$, ][]{Bessell15}, but it shows the  enhancement of some other elements ($\abra{Ca}{H} = -7.26$, $\abra{Mg}{H} = -4.08$, $\abra{C}{H} = -2.55$). 

One plausible explanation for the absence of metal-free star is surface pollution on Pop~III survivors \citep{Yoshii81, Iben83}. It is argued that Pop~III stars can be changed to HMP/UMP stars through pollution by accreting the interstellar medium (ISM) with metals \citep{Shigeyama03, Suda04, Komiya09L}.     
In our previous studies \citep{Komiya10, Komiya15}, we investigate the change of the surface iron abundance  by the ISM accretion for Pop~III survivors,  considering the hierarchical galaxy formation process and the chemical enrichment history. 
We demonstrate that the surface abundance of iron typically amounts to $\feoh = -5$ or $-6$ for Pop~III giants. 
  The predicted distribution of surface metallicity of Pop~III survivors shows good agreement with the observations of HMP/UMP stars including SMSS J0313-6708. 
The ISM accretion on Pop~III stars mainly takes place in mini-halos in which the relative velocity between stars and gas is much lower than in the present MW.

Another possible reason for the absence of the identified metal-free star is the concentration of Pop~III survivors in the MW bulge. 
It is asserted that the very earliest stars in the MW may have centrally concentrated distribution \citep{Diemand05},  
while the  searches for metal-poor stars have targeted  the MW halo. 
Later numerical studies of galaxy formation considering metal enrichment, however, show that Pop~III survivors are likely to be distributed throughout the MW stellar halo \citep{Scannapieco06, Brook07, Gao10}.

In our previous study, we have shown that the predicted surface metallicity distribution of the polluted Pop~III stars well matches the low-metallicity tail of the observed metallicity distribution function \citep[MDF;][]{Komiya15}. 
This result gives a support to our scenario that the observed HMP/UMP stars are the polluted Pop~III stars,   
though is not yet conclusive.   
There are actually other scenarios for HMP/UMP stars as the second or later generations of stars. 
Type~II supernovae (SNe) with very small iron yield (faint SNe) owing to the mixing and fallback mechanism have been argued as plausible progenitors for carbon-rich HMP/UMP stars \citep{Umeda03, Iwamoto05, Tominaga07, Ishigaki14, Marassi14}. 
\citet{Heger10}, \citet{Keller14}, and \citet{Bessell15} proposed SNe II without iron production as progenitors for SMSS 0313-6708 assuming the suppression of the mixing on the basis of the hydrodynamical simulations \citep{Joggerst09}. 
Another possible scenario is that the ejecta from a normal SN II is diluted with very large mass ($\sim 10^8\msun$) to form HMP stars \citep{Karlsson06}. 
In addition,  Meynet(2006) argued that HMP stars are formed from gas ejected by stellar wind from rotating massive stars.

In order to answer whether or not there are low-mass Pop~III survivors, it is crucial to observe stars with truly primordial compositions.  
In this paper, we propose a pathway to observe pure Pop~III survivors, free from surface pollution, by considering an escape of Pop~III survivors from their host mini-halos in which they are born.

Theoretical studies of primordial star formation have shown that Pop~III stars are formed in the dark matter halos of very small mass ($\sim 10^6\msun$, Tegmark et al. 1997, Yoshida et al. 2003). 
Due to the shallow gravitational potential well, low-mass Pop~III stars can escape from these mini-halos to intergalactic space. 
The escaped Pop~III survivors are almost free from the surface pollution by gas accretion because of the very low gas density and low metallicity of the intergalactic space. 
In this paper, we estimate the escape probability and the number of the escaped Pop~III stars around the MW. 
We also investigate the spatial distribution and observability of the escaped Pop~III stars.

Pop~III stars have two possible escape channels from mini-halos as Population~I runaway stars \citep[e.g.][]{Gies86, Fujii11}. 
One is an ejection of the secondary star from a binary system at the supernova explosion of its primary star \citep{Blaauw61}. 
The other is a gravitational slingshot from a star cluster \citep{Poveda67}. 
If Pop~III stars were formed in a star cluster, they have a chance to escape via gravitational, multi-body interactions \citep{Greif11, Stacy13}. 
In this paper, we consider the former channel since we do not have the precise knowledge of the kinematics of stars in Pop~III star clusters.

In this paper, we also estimate the number and distribution of the escaped EMP survivors. 
Observations of these escaped EMP survivors can provide insights into the star-formation history  and merging history of mini-halos.

This paper is organized as follows. 
In the next section, we estimate the escape probability of Pop~III survivors in SN binaries from the mini-halos. 
Then, we compute the number of the escaped Pop~III survivors from the building block mini-halos of  the MW using the hierarchical chemical evolution model based on the extended Press-Schechter (EPS) formalism (Section~\ref{Stree}). 
In Section~\ref{Sdistribution}, we derive the spatial distribution of these mini-halo escapees 
by evaluating the distances between the mini-halos in the EPS merger trees under the spherical collapse approximation. 
We discuss the observability of the escaped Pop~III survivors in Section~\ref{Sobserve} 
and provide a summary in Section~\ref{Ssummary}.

\section{Escape Probability of Pop~III survivors}\label{Sprobability}

We estimate the escape probability of Pop~III survivors from their host mini-halos at the SN explosion in binary systems. 

When the primary star of a binary system explodes as an SN, 
its secondary companion is released from the system
 if more than half of the total mass is lost from the system. 
The released secondary star escapes from the host mini-halo
 if the kinetic energy of the star is larger than the sum of the gravitational potential of the host mini-halo and of the binary system after the SN explosion, i.e.,  

\begin{equation}\label{EqCriterion}
\frac{1}{2} v_{\rm orb}^2 - \frac{G m_{\rm rem}}{a_{\rm bin}} + \phi_{\rm h}(r | M_{\rm h}, z) > 0 ,
\end{equation}  
where $v_{\rm orb}$ is the orbital velocity of the secondary star at the SN explosion, 
 $a_{\rm bin}$ is the separation of a binary, 
 $m_{\rm rem}$ is the mass of the remnant (a neutron star or a black hole) of a primary star, 
 and $\phi_{\rm h}$ is the gravitational potential at a distance, $r$, from the center of a mini-halo with mass, $M_{\rm h}$, and the formation redshift, $z$.

\begin{figure}
\plotone{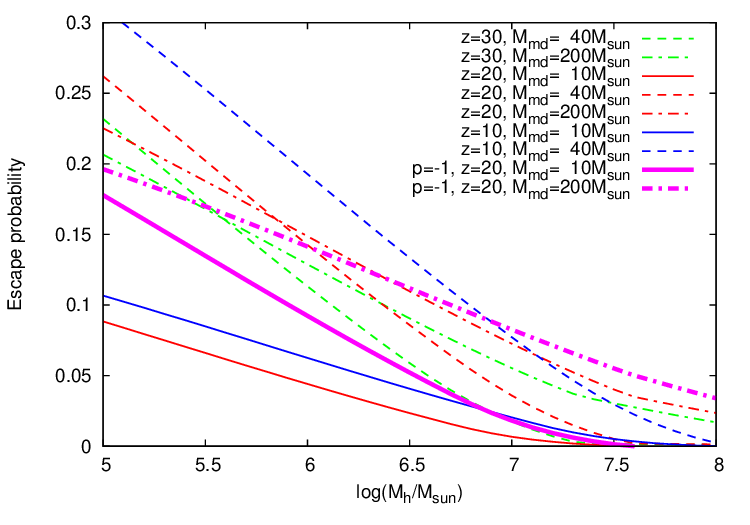}
\caption{
Escape probability of low-mass secondary companions of binaries when their primary stars explode as SNe.  
We plot the escape probability against the mass, $M_{\rm h}$, of their host mini-halos. 
Colors and line types denote the formation redshift, $z$, of mini-halos, the power-law index, $p$, of the mass ratio distributions of binaries ($p=0$ unless otherwise stated), and the median mass, $\mmd$, of the adopted IMFs. 
See the text for details. 
}\label{percent}
\end{figure}

We estimate the escape probability under the following assumptions. 
Pop~III stars are formed at the center of mini-halos. 
The gravitational potential of a mini-halo is given by the NFW profile with the concentration parameter $c=10$. 
The escape velocity at  the mini-halo center is  
 \begin{eqnarray}
 v_{\rm esc} & = & \sqrt{-2 \phi_{\rm h}(0 | M_{\rm h}, z)} \sim 15 \ {\rm km\ s^{-1}} \nonumber \\
 &    \times & \left( \frac{M_{\rm h}}{10^6 h^{-1} \msun} \right)^{1/3} \left(\frac{1+z}{10}\right)^{1/2} \left(\frac{\Omega_{\rm M}(0)\Delta_{\rm c}(z)}{\Omega_{\rm M}(z)18\pi^2}\right)^{1/6} . 
\end{eqnarray} 
 where $\Delta_{\rm c}$ is the overdensity of the virialized halo relative to the critical density of the universe, and $\Omega_{\rm M}(z)$ is the matter density relative to the critical density \citep[e.g.][]{Barkana01, Mo10}.

The escape probability is dependent on the initial mass function (IMF) of Pop~III stars. 
The Pop III IMF is not well understood though it is thought to be more massive than the Pop~I and Pop~II stars. 
 \citet{Hirano14} and \citet{Susa14} followed Pop III star formation in many mini-halos in large-volume cosmological simulations and showed that the mass distribution of Pop III stars ranges from $10\msun$ to $1000 \msun$, and from $1\msun$ to $300\msun$, respectively. 
  On the other hand, \citet{Greif11} and \citet{Stacy13} argued that the mass of Pop III stars shows almost flat distribution between $0.1\msun$ to $10\msun$ and $0.3\msun$ to $30\msun$, respectively. 
 In our previous studies, we have given constraints on the IMF of EMP stars assuming a log-normal function based on the number fraction of carbon-enhanced stars which underwent binary mass transfer events from intermediate-mass AGB stars \citep{Komiya07, Komiya09}, 
 and have yielded the values of the median mass, $\mmdp{EMP}$, values of $2.5 - 20\msun$ and the dispersion, $\sigma$, values of $\sim 0.4$. 
  The rarity of HMP stars and the absence of observed Pop III stars indicate the median mass of Pop III stars, $\mmdp{Pop3}$, should be larger than $\mmdp{EMP}$ \citep{Komiya10, Komiya15}. 
  Here, we use the log-normal IMF with $\sigma = 0.4$ and adopt a wide parameter range for the median mass of $\mmd = 3 - 200\msun$.

On the binary parameters, since there are no observational constraints, we use the same distributions with Pop~I stars as fiducial ones, and adopt other distributions to see the parameter dependence.  
The mass ratio distribution (MRD) is usually described by the power-law $n(q) \propto q^p$, where $q = m_2/m_1$ is the mass ratio between the secondary and primary of the binary system. 
We adopt the power-law index of $p = 0$ or $-1$. 
We note that the most of low-mass stars are born as the secondary member of binary systems under these MRDs with high-mass IMFs. 
For the period distribution, $f(P)$, of binaries, we use  two observational results by \citet[][hereafter DM91]{Duquennoy91} for solar neighbors and \citet[][hereafter Ras10]{Rastegaev10} for Population II halo stars. 
In Ras10, close binaries with short periods are more favored than in DM91. 
The minimum separation of binaries is set at $1000R_{\odot}$ considering the engulfment of the companion star by the massive primary star. 
We assume the circular orbit for simplicity.

The remnant mass, $m_{\rm rem}$, of a primary star is set as a function of its initial mass, $m_{\rm i}$. 
For core collapse SNe of $m_{\rm i} = 8 \mhyph 50\msun$, we adopt 
 the theoretical results for Pop~III SNe by \citet{Woosley95}. 
We assume that $m_{\rm rem} = 0$ for stars of $m_{\rm i} = 140 \mhyph 260\msun$, which explode as pair-instability supernovae \citep[OUSNe;][]{Heger02}.  
Stars of $m_{\rm i} = 50 \mhyph 140\msun$ or $m_{\rm i} > 260 \msun$ are assumed to collapse into blackholes without explosion.

Figure~\ref{percent} shows the escape frequency as functions of mini-halo mass. 
We plot the relative number of the escapees from mini-halos among low-mass ($<0.8\msun$) secondary stars of binary systems. 
In the cases of the very high-mass IMFs with $\mmd \gtrsim 40\msun$, $10 - 20 \%$ of the low-mass stars escape from mini-halos of $10^6 \msun$. 
The escaped stars are the secondary members of close binaries with periods smaller than a few hundred years. 
In the case of $\mmd = 10\msun$, the escape frequency becomes $\sim 3\%$ under the flat MRD and $z = 20$. 
In this case, the majority of Pop~III survivors have intermediate-mass primary companions, and do not escape from binaries.  
The higher $\mmd$ usually increases the escape frequency due to higher orbital velocity. 
In the case of $\mmd = 200\msun$, however, due to the large fraction of primary stars of $m_{\rm i} = 50 \mhyph 140\msun$, the escape probability is smaller than the case of $\mmd = 40\msun$ at $M_{\rm h} \lesssim 10^6\msun$.  
The bottom-heavy MRD of $ p= -1$ predicts a larger escape frequency than the flat one ($p = 0$) because the primary stars tend to be more massive for the secondary stars of a given mass. 
   From more massive mini-halos with $10^8\msun$, the escape frequency is less than a few percent even with the very high-mass IMF due to their deep gravitational potential. 
As a conclusion, the escape fraction in Pop~III survivors is considerable from the low-mass mini-halos, while for the  massive mini-halos, the majority of Pop~III survivors remain in their host mini-halos.

\begin{figure}
\plotone{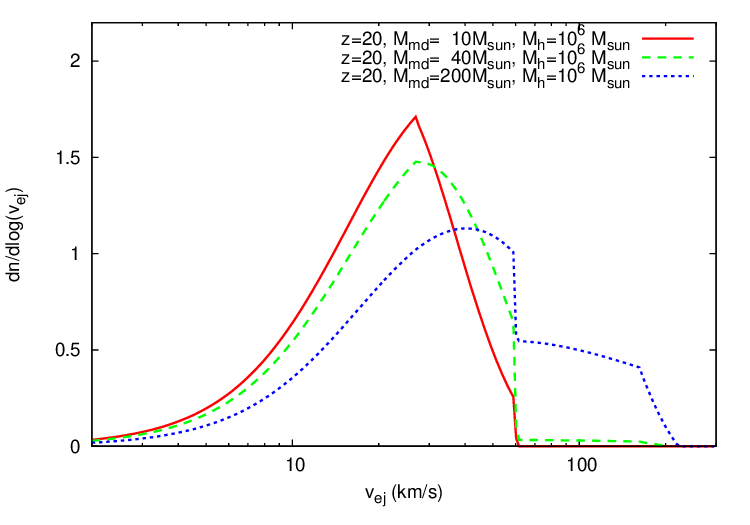}
\caption{
Distribution of the escaping velocity of stars from  mini-halos of $10^6\msun$ at $z=20$. 
The vertical axis is the normalized number frequency. 
Solid, dashed, and dotted lines show the cases of $\mmd = 10\msun$, $40\msun$, and $200\msun$. 
}\label{velocity}
\end{figure}

 After the star escapes from the mini-halo, its velocity approaches to 
\begin{equation}\label{Eq:vej}
v_{\rm ej} = \left( v_{\rm orb}^2 - \frac{2 G m_{\rm rem}}{a_{\rm bin}} + 2\phi_{\rm h}(0 | M_{{\rm h}}, z) \right)^{1/2}.   
\end{equation}
Figure~\ref{velocity} shows the distribution of $v_{\rm ej}$ in the case of $M_{\rm h} = 10^6\msun$ and $z = 20$. 
The typical velocity is slightly larger than the escape velocity of the mini-halo. 
The high-velocity shoulder at $v_{\rm ej} \sim 100 \hbox{ km s}^{-1}$ in the case of $\mmd = 200\msun$ is a contribution from the companions of PISN, which eject all of the mass.
The secondary companion of a PISN has a large escaping velocity because of its large orbital velocity for a given period.

\section{The Number of Escaped Stars}\label{Stree}

We compute the expected total number (Sec.\ref{Snumber})  of the escaped Pop~III and EMP survivors which are formed in the building blocks of the MW.

We use the chemical evolution model with the hierarchical formation process of the MW taken into account \citep{Komiya10, Komiya14, Komiya15}. 
The basic feature and input parameters of the code are described in Sec.~\ref{Smodel}. 
We refer the readers to \citet{Komiya14, Komiya15} for more details about the treatment of chemical evolution.  
In those studies, we have demonstrated that the model  reproduces the MDF and the distribution of $r$-process element abundances of EMP stars. 
In computing chemical evolution, we register all the individual stars and binaries with $Z < 0.1 \Zsun$. 
For each binary system, we set a mass and orbital parameter according to the distribution functions, described above, and 
 compute whether or not the secondary star escapes. 
We describe the results with the fiducial parameter set in Sec.~\ref{Snumber}
 and discuss the parameter dependence in Sec.~\ref{Sparam}.

\subsection{Chemical Evolution Model}\label{Smodel}

We build merger trees  using the method by \citet{SK99} based on the EPS theory. 
The total mass of the merger tree is $M_{\rm MW} = 10^{12}\msun$. 
We consider gas infall according to the dark matter infall rate. 
At $z<10$, we assume that the gas infall rate is zero for mini-halos with virial temperatures of $\Tv<10^4{\rm K}$ because of the cosmic reionization. 

The lower mass limit of mini-halos for star formation is set to give the minimum virial temperature of $T_{\rm vir,min} = 10^3 {\rm K}$ \citep[e.g.][]{Tegmark97, Yoshida03}. 
The star-formation rate, $\psi$, is assumed to be proportional to the gas mass, $\psi=\epsilon_{\star}M_{\rm gas}$,
 where $\epsilon_{\star}$ is the star formation efficiency (SFE) and is set at $\epsilon_{\star} = 10^{-11}{\rm yr}^{-1}$ as a fiducial value.
The mass, $m$, of each metal-poor star is set randomly subject to the IMF with a binary contribution. 
The assumptions about the IMF and binary parameters are the same as described in Section~\ref{Sprobability}. 
We consider negative feedback to the star formation with the following assumptions. 
A massive Pop~III star prohibits star formation in its host mini-halo until it explodes.   
The Lyman--Werner background radiation suppresses Pop~III star formation in the mini-halos of $\Tv < 10^{4}$K formed at $z < z_{\rm LW}$.

The SN yields and explosion energies are taken from \citet{Kobayashi06} and \citet{Umeda02} for SNe II and  for PISNe, respectively. 
We assume that the chemical composition of each mini-halo is homogeneous. 
We consider gas outflow from mini-halos triggered by individual SNe. 
The gas and metal outflow rates are given  as functions of the SN explosion energy and the binding energy of the proto-galaxy. 
The outflow matter from each mini-halo is assumed to  form a galactic wind. 
We follow the evolution of the regions enriched with metals by galactic winds in the intergalactic medium (IGM) assuming the momentum-conserving snowplow shell model in spherical symmetry.

We summarize the ranges and fiducial values of parameters in Table 1. 
We adopt  the parameter ranges as wide as possible since no reliable knowledge is available about the formation rate, the IMF, the binary parameters, and the feedback process of Pop~III stars. 
The values of SFEs are varied from $\epsilon_{\star} = 10^{-12} {\rm yr^{-1}}$ to $10^{-9} {\rm yr^{-1}}$. 
For the Lyman--Werner background, we adopt the parameter range between $z_{\rm LW}=15$ to $30$. 
For the Pop III IMF, we explore the parameter dependence with $\mmdp{Pop3} = 3 - 200 \msun$. 
For EMP stars, we use $\mmdp{EMP}=10\msun$ following our previous studies \citep{Komiya07, Komiya09}.

\begin{table*}
\begin{center}
\caption{Model Parameters}
\label{Tparam}
\scalebox{0.75}{
\begin{tabular}{clccc}
\hline
Parameter  & Description & Fiducial Value & Range & Optimistic Model \\
\hline
$\epsilon_{\star}$	& star formation efficiency	& $10^{-11} {\rm yr^{-1}}$	& $10^{-12}$ -- $10^{-9} {\rm yr^{-1}}$	& $10^{-10} {\rm yr^{-1}}$ \\
$\mmdp{Pop3}$	& median mass of Pop~III stars	& $10\msun$	& $3$ -- $200\msun$	& $10\msun$ \\
$\mmdp{EMP}$	& median mass of EMP stars	& $10\msun$	& $3$ -- $10\msun$ & $10\msun$	\\
$p$ & power-law index of the mass ratio distribution	& 0	& 0 to $-1$ & $-1$ \\
$f(P)$	& period distribution of binaries	& $f_{\rm DM91}(P)$	& $f_{\rm DM91}(P), f_{\rm Ras10}(P)$  & $f_{\rm Ras10}(P)$ 	\\   
$z_{\rm LW}$	& redshift at the Lyman--Werner radiation turns on	& 20	& 15 -- 30	& 15\\
\hline
$M_{\rm MW}$	& total mass of the Milky Way halo	& $10^{12}\msun$	& -	& -	\\ 
 $f_{\rm b}$	& binary fraction	& 0.5	& -	& -	\\
 $T_{\rm vir,min}$	& minimum virial temperature for star formation	& $10^3$ K	& -	& -	\\
 $z_{\rm reion}$	& reionization redshift	& 10	& -	& -	\\
 $Z_{\rm cr}$	& critical metallicity & $10^{-6}\Zsun$ & -	& -	\\
 \hline
\end{tabular}
}
\end{center}
\end{table*}

\subsection{The Escaped Pop~III and EMP Stars}\label{Snumber}

For each low-mass Pop~III and Pop~II binary with $Z<0.1\Zsun$ in the chemical evolution model, 
we compute whether or not the secondary star escapes from its host mini-halo or not by the condition of eq.~(\ref{EqCriterion})
 and count the number of escaping stars. 

\begin{figure}
\plotone{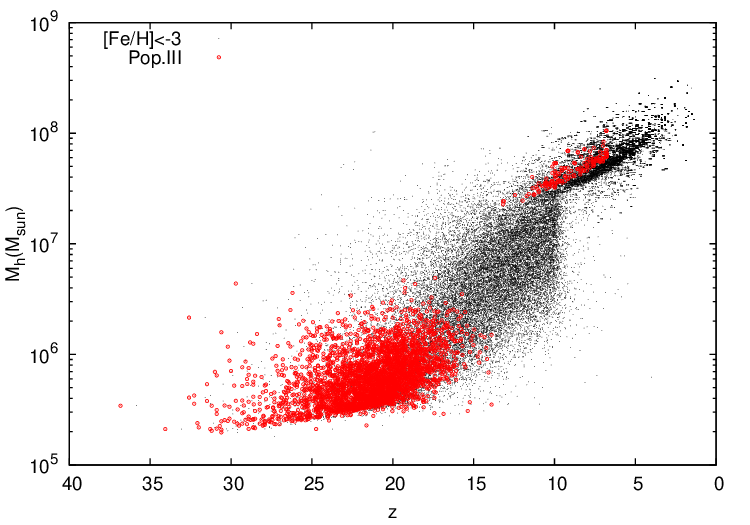}
\caption{
Formation redshift and the mass of host mini-halos for Pop~III (red) and EMP (black) stars in the case of the fiducial parameter set. 
Most of Pop~III stars are formed in small ($\lesssim 10^6\msun$) mini-halos before the Lyman--Werner radiation is turned on at $z = 20$. 
EMP stars are formed as the second or later generations of stars in these mini-halos. 
After $z<10$, the star formation in small mini-halos stops due to the reionization.  
}\label{z-m}
\end{figure}

Figure~\ref{z-m} shows the formation redshift of Pop~III and EMP survivors, and the masses of their host mini-halos in the case of the fiducial parameter set. 
The merger trees with the total masses of $10^{12}\msun$ have more than $2 \times 10^5$ branches with $\Tv>10^3{\rm K}$. 
In the fiducial model, the $\sim 30,000 \mhyph 50,000$ mini-halos formed before $z = z_{\rm LW} = 20$ involve Pop~III stars. 
At $z = 20$, two-thirds of these mini-halos are still metal-free and continue to form Pop III stars. 
Star formation in mini-halos formed at $z<20$ is suppressed due to the Lyman--Werner feedback.  
As shown in Fig.~\ref{z-m}, the vast majority of Pop~III stars are formed in mini-halos with $M_{\rm h} \sim 10^6\msun$ at $z \gtrsim 15$. 
In each star-forming mini-halo, one or two Pop~III star(s) or Pop~III binaries are formed, and $\sim 5\%$ of Pop~III stars have a low-mass secondary companion. 
As a result, the number of low-mass Pop~III survivors is $2800 \mhyph 3800$ in total. 
Among them, 100 - 170 stars escape from their host mini-halos.   
At $z \sim 10$, more massive ($10^7 \mhyph 10^8 \msun$) halos with pristine abundance form $120$ to $220$ low-mass Pop~III stars but their escape probability is very small.

\begin{figure}
\plotone{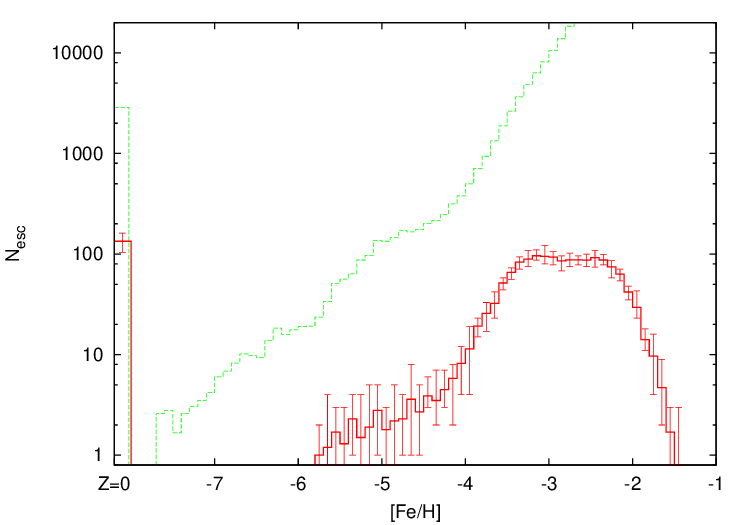}
\caption{
The MDF of the stars that escape from mini-halos as the building blocks of the MW (red solid line). 
The green dashed line denotes the MDF of the whole stars. 
In the fiducial model, 100 -- 170 Pop~III survivors and 1500 or more EMP survivors escape from mini-halos. 
Error bars show the sum of the Poisson error and scatter by 10 computation realizations. 
}\label{exMDF}
\end{figure}

Not only the Pop~III stars but also the second and later generations of stars can escape from mini-halos by the same mechanism. 
Figure~\ref{exMDF} shows the MDF of these mini-halo escapees.  
In the fiducial model, a few percent of the second or later generations of stars with $\feoh < -3$ escape from host mini-halos. 
For $\feoh \gtrsim -3$, the escape probability decreases since the masses of their host mini-halos increase.

\subsection{Parameter Dependence}\label{Sparam}

\begin{figure}
\plotone{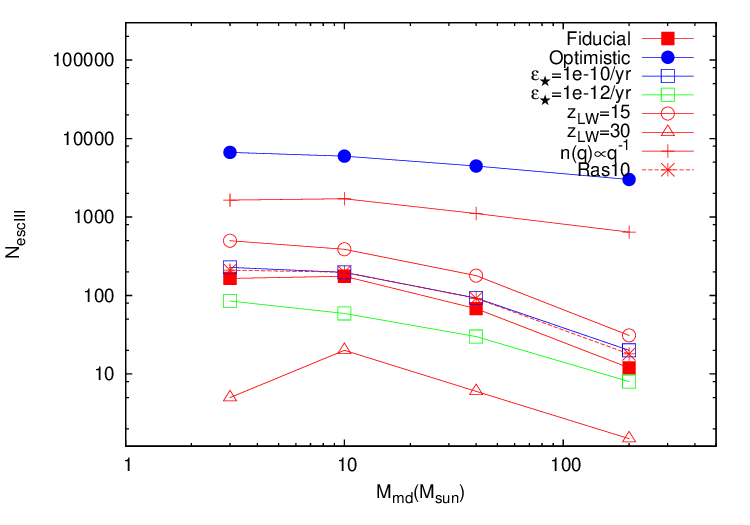}
\caption{
Parameter dependence of the number of escaped Pop~III stars.  
We plot the predicted number against the median mass, $\mmdp{Pop3}$, of the Pop~III IMF. 
Blue, red, and green  colors denote the star formation efficiency, $\epsilon_{\star} = 10^{-10}$, $10^{-11}{\rm yr}^{-1}$, and $ 10^{-12}{\rm yr}^{-1}$, respectively. 
Triangles, squares, and circles show the results of $z_{\rm LW}=15$, 20, and 30, respectively.  
Crosses show the results with the mass ratio distribution of $n(q) \propto q^{-1}$, 
and  asterisks show the results with the binary period distribution by \citet{Rastegaev10}.  
Filled squares and filled circles correspond to the fiducial, $(\epsilon_{\star}, \mmdp{Pop3}, \mmdp{EMP}, p, f(P), z_{\rm LW}) = (10^{-11} {\rm yr}^{-1}, 10\msun, 10\msun, 0, f_{\rm DM91}(P), 20)$, and optimistic, $ (10^{-10} {\rm yr}^{-1}, 10\msun, 10\msun, -1, f_{\rm Ras10}(P), 15)$, parameter sets, respectively (see Table~\ref{Tparam}). 
}\label{param}
\end{figure}

We summarize the parameter dependence of the predicted number, $N_{\rm escIII}$, of the escaped Pop~III stars in Figure~\ref{param}.

As stated above, the fiducial model predicts 2800 - 3800 Pop~III survivors in total, including the escaped  and the bounded ones. 
The number is mainly determined by the IMF and the MRD of Pop~III stars. 
The low median mass ($\mmd$) and the small power-law index ($p$) of the MRD favor low-mass star formation and increase the number of Pop~III survivors.  
On the other hand, high $\mmd$ and small $p$ enhance the escape frequency, as described in Sec.~\ref{Sprobability}.  %

As seen in Fig.~\ref{param}, $N_{\rm escIII}$ decreases as $\mmd$ increases at $\mmd > 10 \msun$ because the number of low-mass stars decreases. 
On the other hand,  the number is almost independent of $\mmd$ at $\mmd \leq 10\msun$ because the number of low-mass stars increases as $\mmd$ decreases, but the number of massive primary companion stars decreases. 

In the case of $p = -1$, the escape fraction is $\sim 10$ times larger than in the case of $p = 0$.   
With the period distribution of Ras10, the value of $N_{\rm escIII}$ becomes $\sim 1.5$ times larger compared with DM91 due to the large frequency of close binaries with high orbital velocity. 
 The number of the escaped stars is simply proportional to the binary fraction if the number of systems is fixed.

The number of Pop~III stars and the escape frequency are also dependent on $z_{\rm LW}$ and $\epsilon_{\star}$. 
The high SFE makes $N_{\rm escIII}$ large since many stars are formed before mini-halos grow in mass.  
If the Lyman--Werner feedback takes place at very high redshift, e.g., $z=30$, few Pop~III survivors  escape since the star formation is suppressed in most of low-mass mini-halos with $\Tv < 10^4$ K.

Under an optimistic parameter set as shown in Table~1, approximately $10,000$ Pop~III survivors escape from the building blocks of the MW.

The number, $N_{\rm escEMP}$, of escaped EMP stars shows the similar parameter dependence. 
We note that the number of escaped Pop~III stars is almost independent of the parameters of EMP stars.

\section{Spatial Distribution of the Escaped Stars}\label{Sdistribution}

\subsection{Model and Assumptions}

We first specify the spatial distribution of mini-halos in the EPS merger trees under the spherical collapse approximation. 
We then compute the orbit of each escaped star formed in the chemical evolution model and estimate the distribution of escaped stars at $z = 0$. 
 In this study, we only deal with the evolution of the Galactocentric distance of stars.

The radial component of the equation of motion for the mini-halo escapees is written as follows,
\begin{equation}
\frac{d^2 r_{*}}{dt^2} = -\frac{G M_{\rm in}(r_{*},t)}{r_{*}^2} + \frac{\Lambda c^2}{3} r_* + \frac{l_{*}^2}{r_{*}^3} , 
\end{equation}
 where $r_{*}$ is the Galactocentric distance\footnote{$r_{*}$ is the distance in a fixed (not comoving) coordinate frame} of an escaped star, 
 $M_{\rm in}(r_{*}, t)$ is the total mass inside $r_{*}$ at time $t$, 
 $\Lambda$ is the cosmological term, 
 and $l_{*}$ is the specific angular momentum of the star.

We assume that a main branch of the merger tree (proto-MW) stays at the position where the MW is now located. 
We take the initial distance when a star escapes from a mini-halo to be the distance between the mini-halo and the proto-MW. 
We estimate the distance, $r_n$, between the $n$-th mini-halo and the proto-MW based on the EPS merger tree as follows \citep{Komiya15}.  
The Press--Schechter formalism considers the formation of a halo by assuming the spherical collapse of a density perturbation. 
The evolution of the radius of the uniformly overdense spherical region with mass $M$ which collapses at $t_{\rm c}$ is given by integrating the following equation of motion,  
\begin{equation}\label{eq:rshell}
\frac{d^2 r_{}(t)}{dt^2} = -\frac{G M}{r_{}} + \frac{\Lambda c^2}{3} r_{} ,
\end{equation} 
 with the initial and collapse  conditions of $r(0) = r_{}(t_{\rm c})=0$. 
  As a corollary, the radius can be described as a function of $t$, $M$, and $t_{\rm c}$, 
\begin{equation}\label{eq:rtMtc} 
r = r(t | M, t_{\rm c}). 
\end{equation}
The EPS formalism considers the ``merger" of two halos as a gravitational collapse of a region incorporating the two halos. 
In this framework, the distance, $r_n$, of a mini-halo which merges with the proto-MW at time $t_{{\rm merge}, n}$ 
 may well be approximated to the radius, $r$, of the region of the mass, $M_{0}(t_{{\rm merge}, n})$,  of the proto-MW at the time $t_{{\rm merge}, n}$. 
In other words, we approximate the Galactocentric distance, $r_n$, of the $n$-th mini-halo as   
\begin{equation}\label{eq:rhalon}
r_n(t) = r(t | M_{0}(t_{{\rm merge}, n}), t_{{\rm merge}, n}) ,
\end{equation}
   with $M_0(t)$ and $t_{{\rm merge}, n}$ are given by a merger tree. 

The typical value of $r_{n}(t_{\rm e})$ is around $50 - 100$kpc for the progenitor mini-halos of the escaped Pop~III stars.

The initial velocity of an escapee relative to the proto-MW is given by the sum of 
   the velocity of the escapee, $v_{\rm ej}$, relative to its host mini-halo in eq.~(\ref{Eq:vej}) and the velocity of the host mini-halo relative to the proto-MW at the time, $t_{\rm e}$, when the star escapes. 
   The radial component is written as 
\begin{equation}\label{EqInitialV}
v_{*}(t_{\rm e}) = v_{\rm ej} \cos\theta + \left.\frac{dr_{n}(t)}{dt}\right|_{t_{\rm e}} ,
\end{equation}
where  
 $\theta$ is the angle of the escapee velocity to the radial direction and is set random in the following.

The specific angular momentum of a star is set at 
\begin{equation}
l_* = r_{n}(t_{\rm e}) v_{\rm ej} \sin\theta .  
\end{equation}
The escaping velocities of Pop~III stars are around $v_{\rm ej} \sim 30 \hbox{ km s}^{-1}$, as shown in Fig.~\ref{velocity}.  
The averaged specific angular momentum of Pop~III survivors is $\sim1200 \hbox{kpc} \cdot \hbox{km s}^{-1}$ in the fiducial model. 
We neglect the tangential proper motion of mini-halos. A numerical study shows that dark matter sub-halos have specific angular momentum of $0.51 R_{\rm vir} V_c$ on average, where $R_{\rm vir}$ is a virial radius and $V_c$ is the circular velocity of a primary halo when a sub-halo falls into the virial radius \citep{Jiang08}. 
Given that the angular momentum grows approximately linearly with time until a mini-halo separates from the Hubble flow \citep[e.g.][]{Peacock99},  
the angular momentum of mini-halos at $t_{\rm e}$ is smaller ($\sim 500 \hbox{ kpc} \cdot \hbox{km s}^{-1}$) than $r_n v_{\rm ej} \sin\theta$ for the majority of the escaped Pop~III stars due to high redshift when Pop~III stars escape.   

We also neglect the change of radial velocity and angular momentum of the escaped stars by encounters with satellite halos since the event rate for close encounters with mini-halos other than the proto-MW is smaller than unity, as shown in the Appendix.

For the mass, $M_{\rm in}(r_{*}, t)$, inside $r_*$, we use the following equation,  
\begin{equation} 
M_{\rm in}(r_*, t) = \begin{cases}
		M_{0}(t)  \frac{\log(1+r_* c/R_{{\rm vir}, 0}) - r_*/(R_{{\rm vir}, 0}/c+r_*)}{\log(1+c)-c/(1+c)} 
\\ \qquad\qquad\qquad \hbox{ for } r_* < R_{{\rm vir}, 0}(t) \\ 
M_{0}(t_{\rm c}(r_*, t)) 					
 \quad \hbox{ for } R_{{\rm vir}, 0}(t) < r_* < r_{\rm MW}(t)   \\ 
		M_{\rm MW} + \frac{4\pi}{3}(r_*^3 - r_{\rm MW}(t)^3)\rho_{\rm av}
\\ \qquad\qquad\qquad \hbox{ for } r_{\rm MW}(t) < r_* ,
		\end{cases} 
\end{equation}
 where $R_{{\rm vir}, 0}(t)$ is the virial radius of the proto-MW  and  
 $r_{\rm MW}(t)$ is the radius of the overdense region which collapses to the present MW. 
 Inside the virial radius $r_* < R_{{\rm vir}, 0}(t)$, we adopt the NFW density profile. 
We  put $r_{\rm MW}(t)$ equal to $r(t | M_{\rm MW}, t_{\rm univ})$ in eq.~(\ref{eq:rtMtc}) with $t_{\rm univ}$, the age of the universe at $z = 0$.  
Outside this radius, $r_{\rm MW}(t)< r_*$, we assume the average density, $\rho_{\rm av}$, of the universe.
At the intermediate distance range between $R_{{\rm vir}, 0}(t)$ and  $r_{\rm MW}(t)$, $M_{\rm in}$ is the sum of the mass of all the mini-halos with $r < r_*$ in a merger tree and IGM inside $r_*$. 
We estimate the mass from eq.~(\ref{eq:rshell}) and the merger tree as follows.  
A mass element at distance $r$ at time $t$ moves following eq.(\ref{eq:rshell}) and accretes to the proto-MW at $t_{\rm c}$. 
Mass inside this orbit is $M_0(t_{\rm c})$. 
In other words, we can give $t_{\rm c}(r, t)$ as a function of $r$ and $t$ by solving {$r = r(t | M_0(t_{\rm c}), t_c)$} for $t_{\rm c}$ numerically. 
Under the spherical collapse approximation, the mini-halos and IGM inside $r$ at $t$ accrete to proto-MW before $t_c(r, t)$. 
Therefore, we can approximate the total mass inside $r_*$ at $t$ as the mass of a proto-MW at $t_{\rm c}$, i.e., $M_{0}(t_{\rm c}(r_*, t))$.

\begin{figure}
\plotone{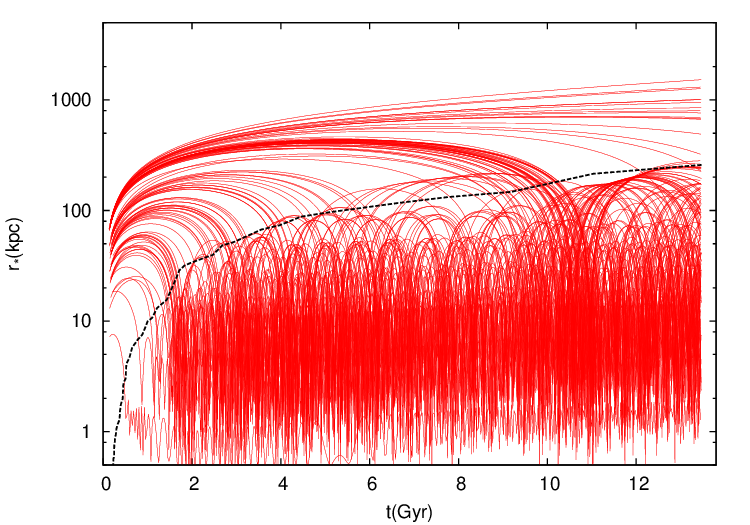}
\caption{Orbits of the escaped Pop~III stars for a computation realization with the fiducial parameter set. 
 The Galactocentric distance is plotted as functions of time. 
Black dashed line denotes the virial radius of the main branch of the merger tree (proto-MW). 
}\label{orbit}
\end{figure}

\subsection{Results}
 
\begin{figure}
\plotone{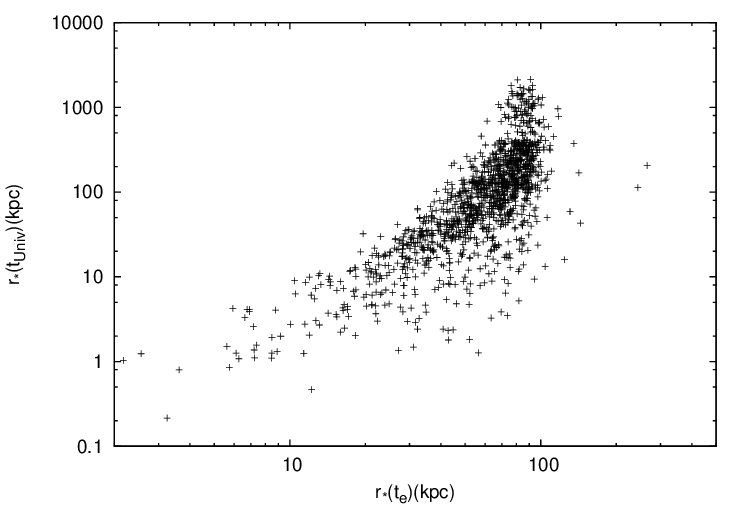}
\plotone{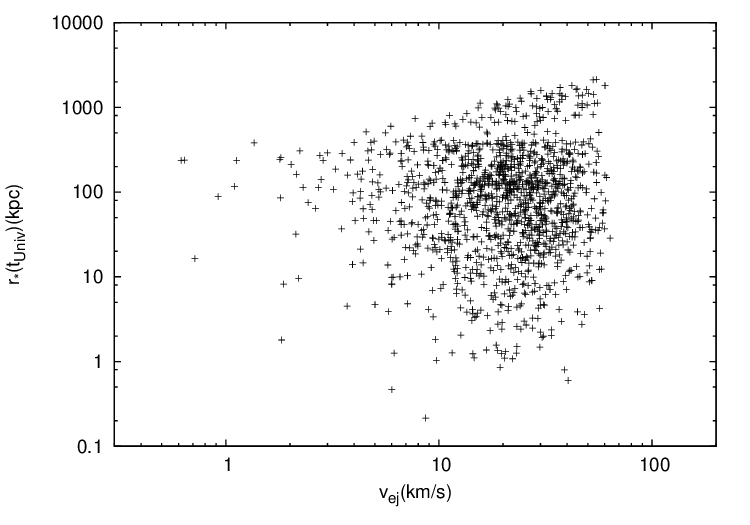}
\caption{
Galactocentric distance of the escaped Pop~III stars at $z=0$  
 against the positions of their original host mini-halos when they escape (top panel),  
 and against the ejection velocity relative to their host mini-halos (bottom panel) 
for the fiducial model.  
}\label{rfin}
\end{figure}

\begin{figure}
\plotone{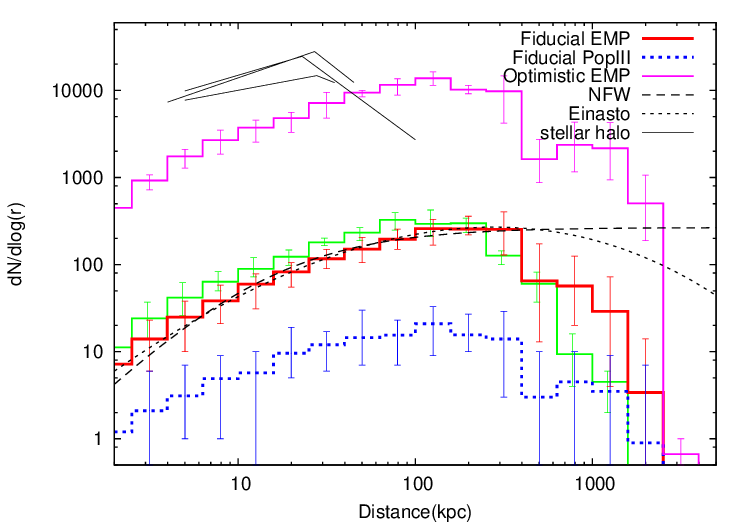}
\caption{
Predicted distributions of the Galactocentric distance of all the mini-halo escapees (solid red histogram) and the escaped Pop~III stars (dashed bflue {histogram}) for the fiducial model. 
The magenta histogram shows the number of mini-halo escapees in each spherical shell in the case of the optimistic model. 
Error bars show the maximum and minimum values for 10 realizations. 
The thin green histogram denotes the results with merger trees without a major merger at low redshift. 
Dashed and dotted black lines are the NFW profile and the Einasto profile, respectively, fitted to the red histogram. 
Solid black lines on the top show the broken power-law fit to the observed density profile of the stellar halo, which declines with a power-law index of $-3.8$ \citep{Sesar11}, $-4.5$ \citep{Watkins09}, or $-4.6$ \citep{Deason11} beyond $25 - 30$ kpc. 
}\label{dist}
\end{figure}

We follow the motion of all mini-halo escapees from $t_{\rm e}$ to $t_{\rm univ}$. 
Figure~\ref{orbit} shows the variation of the radial distance  of the escaped Pop~III stars in one computation{al} realization with the fiducial parameter set. 
A majority of escapees fall into the MW halo at a few to 10 Gyrs after the escape.

The progenitor mini-halos of the MW are distributed in $\sim 80$kpc at $z = 20$ and $\sim 150$ kpc at $z = 10$. 
The velocity of these mini-halos by Hubble flow is $\sim 300 \hbox{ km s}^{-1}$ at $z = 20$ and $\sim 200 \hbox{ km s}^{-1}$ at $z=10$. 
On the other hand, the escaping velocity ($v_{\rm ej}$) of stars from mini-halos is $v_{\rm ej} \lesssim 60 \hbox{ km s}^{-1}$, as shown in Figure~\ref{velocity}.  
Therefore, most of the ``escaped" stars cannot get the enough velocity to escape from the MW. 
The cosmological term is negligible.

Figure~\ref{rfin} shows the distributions of the escaped Pop~III stars on the $r_*(t_{\rm e})$-$r_*(t_{\rm univ})$ plane (top panel) and the $v_{\rm ej}$-$r_*(t_{\rm univ})$ plane. 
There is a clear gap at $r_*(t_{\rm univ}) \sim 300$--$600$ kpc.  
 Stars that fell onto the MW halo are distributed inside or slightly outside the virial radius ($260$ kpc) of the MW halo. 
For the stars at $r_*(t_{\rm univ})< 300$ kpc, we see a clear positive correlation of the present radial distance with the initial radial diatance (top panel), but no correlation with the escaping velocity (bottom panel).  
The motion of these stars is mainly dominated by the Hubble flow and the gravity of the proto-MW. 
On the other hand, for stars with $r_*(t_{\rm univ}) \gtrsim 600$ kpc, the present distance, $r_*(t_{\rm univ})$, is dependent on $v_{\rm ej}$.  
Their host mini-halos were distant from the proto-MW when they escaped. 
These stars are not bounded to the MW and will be floating in the intergalactic space with nearly constant velocity.

Figure~\ref{dist} shows the predicted spatial distribution of the escaped Pop~III survivors (dashed blue lines) and all the mini-halo escapees (solid red line) at the present for the fiducial model. 
The distributions are peaked at $\sim 100$ kpc, the outer region  of the MW's dark halo. 
20--30\% of escapees spread further outward in the intergalactic space outside the MW, and  
 are distributed beyond  $\sim 1$ Mpc away from the MW.

We can fit the distribution  of escaped stars  by the NFW profile or the Einasto profile.     
  In Fugure~\ref{dist}, we plot the NFW profile with the scale radius $r_s = 11$kpc and the Einasto profile with the shape parameter $\alpha = 0.16$ and $r_s = 11$kpc.  
These parameter values are comparable with those of the MW mass halos in numerical simulations \citep[e.g.,][]{Tissera10}.
The distribution of all the escaped stars well agrees with these profiles at the radius $r \lesssim 300$ kpc, while it grows much smaller beyond this radius.

In comparison, the Pop III escapees exhibit slightly more centrally condensed distribution than the EMP escapees.   
   This may stem from the difference in the ratio of $v_{\rm ej}$ to the velocity of the Hubble flow; 
   it is larger for the EMP escapees because of the decrease in the Hubble flow, which renders the spread of initial velocity larger.

The magenta line denotes the distribution under the optimistic parameter set. 
The distance distribution is almost independent of the IMF, the binary parameters, and the SFE, while the number of escaped Pop~III stars is dependent on them.

The predicted distribution is slightly dependent on the merging history of a galaxy. 
Galaxy formation studies indicate that disk-dominated galaxies  such as the MW experienced no major mergers at low redshift. 
 We select  merger trees with $M_{0}(z=2) > 0.75 \times M_{\rm MW}$, and show the result with these trees as a green line in Figure~\ref{dist}.    
In this case, the distribution of mini-halos is more centrally concentrated at the formation epoch of Pop~III or EMP stars. 
The predicted distribution of the escaped stars at $z=0$ is also slightly shifted toward the MW center.  
In particular, only 6]-10 \% of the escaped stars is outside the virial radius and the number of stars at $r_* > 600$ kpc is very small.

\section{Observability}\label{Sobserve} 
In this section, we compare the predicted distribution of escaped Pop~III stars with the distribution of field halo stars, 
 and discuss a strategy to search for the escaped Pop~III stars.

The number frequency of the escapees among the Pop~III stars is $\sim 3\%$ in the fiducial model,  as stated in Sec.~\ref{Snumber}. 
This frequency is dependent on the Galactocentric distance. 
As stated above, the predicted distribution of mini-halo escapees is similar to the dark matter halo and more extended than the stellar halo. 
The observed density profile of the MW stellar halo at the solar vicinity is fitted by $\rho \propto r^{-3}$. 
Recent observations have found that the stellar density declines more rapidly beyond $25 - 30$ kpc with a power-law index of $-3.8$ \citep{Sesar11}, $-4.5$ \citep{Watkins09}, or $-4.6$ \citep{Deason11}, as shown in thin solid black lines in Figure~\ref{dist}.   
Numerical studies for the formation of a stellar halo also predict a power-law index with $-3$ to $-4$ \citep{Bullock05, Font11}.

\begin{figure}
\plotone{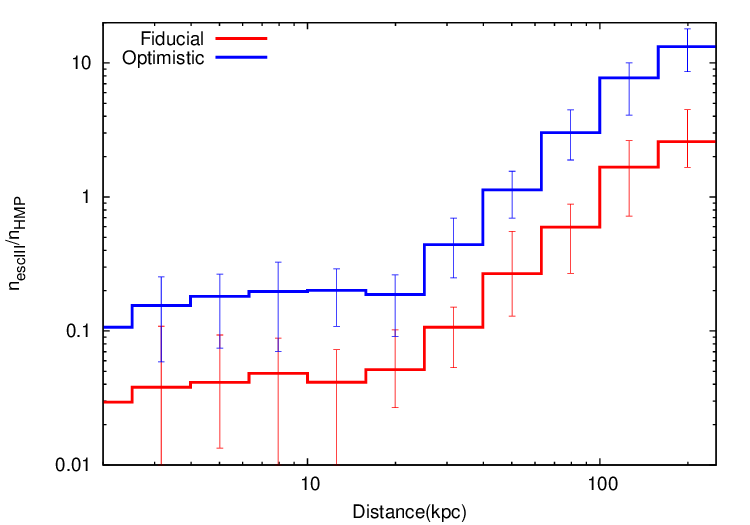}
\caption{
Predicted number ratio of escaped Pop~III stars to HMP stars against the Galactocentric distance. 
Red and blue lines show the results of the fiducial model and the optimistic model, respectively. 
For the total number of HMP stars, 
we adopt the result of \citet{Komiya15} under the polluted Pop~III star scenario for HMP stars. 
The spatial distribution of the HMP stars is assumed to be the same with the stellar halo. 
}\label{dist3}
\end{figure}

In Figure~\ref{dist3}, we plot the predicted number ratio between the escaped Pop~III stars and  the HMP stars against the Galactocentric distance. 
As mentioned above, we assume that the  Pop~III stars that remain in the host mini-halos undergo the surface pollution to be observed as HMP stars \citep{Komiya09, Komiya15}. 
It is shown that one-third of the polluted Pop~III stars are distributed at $\feoh < -5$ in our model. 
For the spatial distribution  of HMP stars, we adopt the density profile of the MW stellar halo derived by \citet{Watkins09} from the observation of RR Lyrae stars;  
\begin{equation}
n(r) \propto 
\begin{cases} 
r^{-2.4} & (r < 23{\rm kpc}) \\
r^{-4.5} & (r > 23{\rm kpc}) .
\end{cases}
\end{equation}
We extrapolate this power-law profile from the MW center to the virial radius since  their observational sample is limited to the distance of $5 < r < 115$ kpc.

The red and blue lines show the result for the fiducial model  and the optimistic model, respectively. 
At $r \lesssim 20$ kpc, the number ratio between the escaped Pop~III stars and the HMP stars is $\sim 1/25$ for the fiducial case and $\sim 1/5$ for the optimistic case. 
The frequency of escaped Pop~III stars increases in the outer halo. 
At $r \simeq 100$ kpc, the expected number of escaped Pop~III stars grows comparable to that of HMP stars.  
Observationally, two HMP stars are found in the sample of the HES survey, and three additional HMP stars are discovered by later observations. 
The absence of metal-free stars in the current samples is consistent with the predicted low frequency of escaped Pop~III survivors among nearby stars.  
Our result indicates that a survey of the outskirt of the MW halo or a survey of nearby halo stars with a 10 times larger sample volume than the HES survey may find pristine Pop~III star(s).

\begin{figure}
\plotone{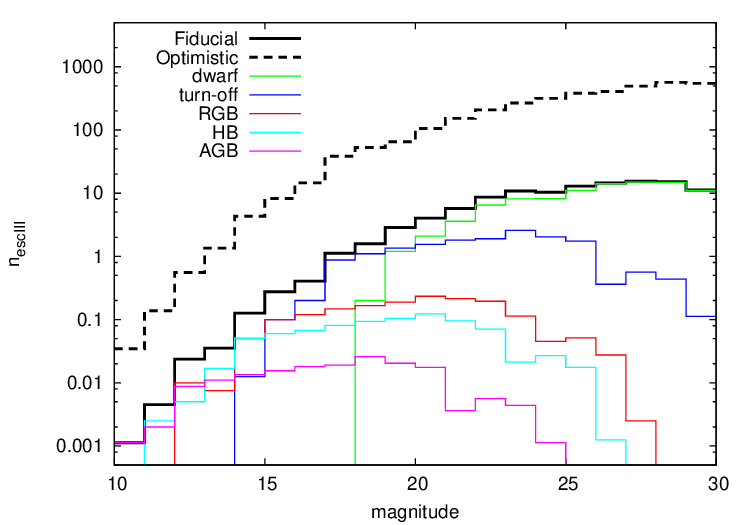}
\caption{
Predicted magnitude distribution of the escaped Pop~III stars around the Milky Way. 
Colored lines denote the magnitude distributions of main-sequence stars (green), the subgiant stars (blue), the red giant stars (red), the horizontal branch stars (cyan), and the AGB stars (magenta). 
Solid and dashed black lines denote the total numbers for  the fiducial and the optimistic parameter sets, respectively. 
}\label{magnitude}
\end{figure}

In Figure \ref{magnitude}, we give the predicted apparent magnitude distribution of the escaped Pop~III survivors. 
The solid and dashed black lines correspond to the results of the fiducial and the optimistic models, respectively. 
The colored lines show the number of stars at different evolutionally stages in the fiducial model. 
We assume the averaged absolute magnitudes of $M_V = 3.0, 0.5, -0.3$ and $-2.0$, respectively, for the Pop~III survivors during the subgiant, the red giant, the horizontal branch and the asymptotic giant branch (AGB) phases  
 based on the stellar evolution calculation of Pop~III stars with $0.8\msun$ \citep{Suda10}. 
The duration of each evolutionary phase is taken into consideration as the corresponding mass ranges of $\Delta m = 0.1\msun, 0.01\msun, 0.005\msun$, and $0.001\msun$, respectively.  
For main-sequence stars, we adopt the mass-luminosity relation $L = (m/\msun)^{3.5} \times L_{\odot}$. 
   We set the Galactocentric distance of the solar system at 8kpc. The figure indicates that the majority of escaped Pop~III stars in the RGB phase have apparent magnitudes brighter than $V \sim 20$mag.

If we adopt $4 \times 10^8\msun$ for the total mass of the MW stellar halo \citep{Bell08} and the IMF of \citet{Chabrier03}, 
 the number ratio between the escaped Pop~III stars and the field halo stars is 1/1,000,000 in the fiducial case and 1/30,000 in the optimistic case.  
 If we observe RGB stars at $M_V = 19 - 21$, however, the number fraction of the escaped Pop~III stars among the halo stars amounts to $\sim 1/40,000$ for the fiducial model and $\sim 1/1000$ for the optimistic model. 
A multi-object spectrograph can be a useful tool to select candidates of the mini-halo escapees. 
For example,  the Prime Focus Spectrograph as a next generation instrument of the Subaru telescope will carry out the spectroscopy of 2400 targets within 1.3 degree diameter. 
It may reach to $V = 21$ mag by 2 hr exposure with the resolution $R = 3000$. 
We can select red giant EMP star candidates from the spectroscopic data provided by such an instrument.

\begin{figure}
\plotone{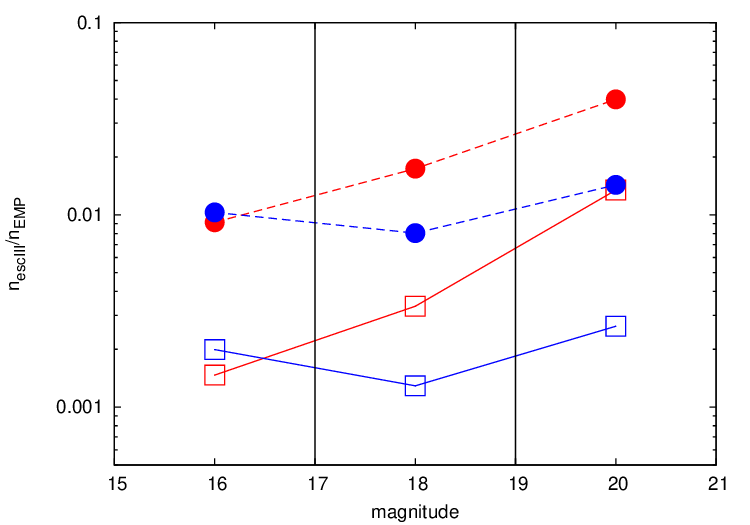}
\caption{
Number fraction of the escaped Pop~III stars among stars with $\feoh<-3$ for red giants (red) or subgiant stars (blue).  
We plot the fraction for three apparent magnitude ranges of $V=15$ - 17 mag, 17 - 19 mag, and 19 - 21 mag. 
Open squares and filled circles show results for the fiducial and optimistic parameter sets, respectively. 
}\label{detection}
\end{figure}

In order to sort out  the EMP stars and the Pop~III stars from candidate stars, high-resolution spectroscopic observations are required. 
Figure~\ref{detection} shows the number fraction of escaped Pop~III survivors among stars with $\feoh < -3$ as a function of the apparent magnitude. 
We plot the number fraction among red giants (red) and the subgiant stars (blue) in our fiducial (open square) and optimistic (filled circle) models. 
About 1\% of red giant stars with 19 -- 21 mag are the escaped Pop~III stars for the fiducial model. 
For  the optimistic model, the majority of RGB stars with $19 - 21$ mag are the mini-halo escapees, and $\sim 4\%$ of them are escaped Pop~III stars.  
The reason for the depression at 17 -- 19 mag for the subgiant stars in this figure is  that stars around the MW center fall in this magnitude range. 
If we target stars in the high galactic latitude area, the fraction of Pop~III stars  is almost flat against magnitude at $\lesssim 20$mag for the subgiant stars since the number fraction against distance is almost flat at $\sim 20$ kpc, as shown in Fig.~\ref{dist3}.

This figure indicates that the follow up observation with high-resolution spectroscopy for several dozens of RGB stars with $\feoh<-3$ at a magnitude range of $V = 19 - 21$ gives us a chance to find a few escaped Pop~III survivors. 
These faint Pop~III or EMP stars can be targets for a high-resolution spectroscopy using the Thirty-Meter Telescope (TMT) or the European Extremely-Large Telescope (E-ELT).

We note that these Pop~III stars show the enhancements of carbon and nitrogen abundances, and also of oxygen abundance but to a much lesser degree, during the horizontal branch and the AGB phases. 
Low-mass Pop~III stars undergo the proton ignition into the helium convection during the core helium flash, ignited at the tip of RGB \citep{Fujimoto90}, and helium-flash-driven deep mixing occurs to dredge up the nuclear products of helium flash  to their surfaces \citep{Hollowell90,Fujimoto00,Suda04, Suda10, Campbell08, Campbell10}. 
It may also enhance $s$-process elements \citep{Fujimoto00,Iwamoto04}.  
On the other hand, no mechanisms are known to enhance iron group elements in Pop~III escapees.

As shown in Figure~\ref{dist}, we predict that some stars are expelled into the intergalactic space at a distance of $\sim 1$ Mpc. 
   Accordingly, we may also observe the Pop~III and EMP stars that are ejected into intergalactic space from the Andromeda galaxy in the same way around the MW. 
   When we assume that the mass of the Andromeda galaxy is $1.5 \times 10^{12}\msun$,   
the expected number of stars which come to the MW is $1.5 N_{\rm escEMP} (\pi R_{\rm vir,0}^2(t_{\rm univ})/4 \pi d^2) \sim 10$ in the fiducial case and $\sim 700$ in the optimistic case, where $d = 780$kpc is the distance of the Andromeda galaxy.   
  These stars becomes hypervelocity stars with the velocities exceeding the escape velocity, and have a possibility of being observed by the Gaia satellite.

   The existence of escaped stars in intergalactic space may also support the hierarchical formation scenario of galaxies. 
In addition, the number ratio between Pop~III stars inside and outside the MW halo can provide a constraint on the merging history of the MW, as shown in Figure~\ref{dist}. 
It is presently difficult, however, to find such stars since they are very faint even at the giant phases ($\sim 26$ mag for RGB and $\sim 23$ mag for AGB).

\section{Summary and Discussion}\label{Ssummary}

In this paper, we have estimated the escape probability of low-mass Pop~III stars from mini-halos in which they are formed.  
Since Pop~III stars are formed in halos with very small masses ($\sim 10^6\msun$), a few to 20\%, depending on model parameters, of low-mass Pop~III stars in binaries are expected to escape when their primaries explode as SNe. 
While surfaces of Pop~III survivors that remain bounded in mini-halos can be polluted by accreting surrounding gas with metals, the escaped Pop~III stars can keep the surface  in the pristine chemical composition.

 From the mini-halos as building blocks of the Milky Way (MW), a few hundred Pop~III survivors can escape if we adopt the high-mass IMF and assume the same binary parameter distribution as Population I stars.  
The number of such stars depends on the IMF, the MRD, the star-formation rate, and the efficiency of the negative feedback on Pop~III star formation. 
In the case of an optimistic parameter set, we predict tens of thousands of the escaped Pop~III stars.

We also compute the orbits of escaped Pop~III stars to see their spatial distribution.   
Most of the escaped stars fall back onto the dark matter halo of the MW and are distributed around $100$ kpc away from the MW center. 
The predicted distribution profile is similar to the density profile of the dark matter halo.

It is conceivable that the surface of an escaped Pop~III star is polluted by ejecta from a SN explosion of its primary companion. 
However, a numerical simulation of the interaction between SN ejecta and a companion star shows that SN ejecta is not retained in the secondary star since the surface of the secondary star is stripped by shock heating \citep[e.g.][]{Hirai14}.

A gravitational slingshot in star clusters is another mechanism by which to escape from mini-halos, which is not considered in this work. 
Recent numerical simulations of Pop~III star formation show that a half of low-mass Pop~III stars can escape from their host mini-halos with velocities of several to hundreds of $\hbox{km s}^{-1}$ through gravitational multi-body interaction \citep{Greif11, Stacy13}. 
The expected number of the escaped Pop~III stars in this channel can be similar to or larger than our optimistic model.    
On the other hand, the distance and magnitude distribution of the escaped Pop~III stars are expected to be similar in both channels since the velocity of the escapees by gravitational slingshot is similar to the escaped stars from SN binaries.

We have a chance to observe escaped Pop~III stars by a large-scale survey of giant stars in the outskirts of the MW.   
The predicted number density of escaped Pop~III stars is comparable to HMP stars at a distance of $\sim 100 {\rm kpc}$, the latter of which remain in their host mini-halos and are subject to the surface pollution by the accretion of interstellar gas. 
We expect that the number fraction of the escaped Pop~III stars becomes 1/1000 -- 1/40,000 among halo RGB stars in the range of 19 -- 21 mag. 
Multi-object spectrographs such as the Subaru Prime Focus Spectrograph can be a useful tool for selecting extremely metal-poor objects as candidates of Pop~III stars. 
Follow up observations of high-resolution spectroscopy  for several dozens of EMP stars in the outskirts of the MW using the TMT or E-ELT will provide a chance to identify Pop~III survivors.

Observation of stars with zero surface metallicity would be a smoking gun for low-mass star formation in a zero-metallicity environment. 
The existence of such low-mass Pop~III stars supports the scenario that HMP stars are polluted Pop~III stars. 
If Pop~III stars are mainly distributed in the very outer region of the MW, it  indicates that HMP stars are polluted counterparts in the stellar halo of the MW.

\appendix

\section{Encounter with Satellite Halos}

In section~\ref{Sdistribution}, we neglect the contribution from mini-halos other than the main branch of the merger tree in computing the orbits of the escapees from mini-halos.  
Here, we show that it is negligible that the escaped Pop~III stars are trapped or scattered by other mini-halos other than the proto-MW. 

The orbit of an escaped star is significantly changed by a mini-halo if the gravitational potential of the mini-halo is larger than the kinetic energy, 
 i.e., the impact parameter, $b$, is smaller than $b_{\rm m}$ as defined by the following equation 
\begin{equation}
 | \phi_{\rm h}(b_{\rm m} | M_{\rm h}, z) | = \frac{1}{2} v_{\rm rel}^2 ,
\end{equation} 
 where $\phi_{\rm h}(r | M_{\rm h}, z)$ is the gravitational potential at a distance $r$ from the center of a halo at redshift $z$ with mass $M_{\rm h}$, 
 and $v_{\rm rel}$ is the relative velocity of a star and the mini-halo. 
For the gravitational potential, we use the NFW profile at $r \leq  r_{\rm vir}$ and $\phi_{\rm h} = GM_{\rm h}/r$ at $r > r_{\rm vir}$, where $r_{\rm vir}$ is the virial radius of the mini-halo. 
In the case of mini-halo escapees, $v_{\rm rel}$ is similar to $v_{\rm ej}$.  The cross section of the close encounter is $\sigma = \pi b_{\rm m}^2 (z(t), M_{\rm h}, v_{\rm rel})$.

The frequency of encounter is described as follows 
\begin{equation}
\int^{t_{\rm uni}}_{t_{\rm e}} dt \int^{M_{\rm max}}_{M_{\rm min}} dM_{\rm h} \frac{dn}{dM_{\rm h}}(M_{\rm h}, z(t))  \sigma(t, M_{\rm h}, v_{\rm rel}) v_{\rm rel} , 
\end{equation}
 where $dn/dM_{\rm h}$ is the number density of halos of mass between $M_{\rm h}$ and $M_{\rm h}+dM_{\rm h}$, and given by the Press--Schechter theory. 
We adopt $t_{\rm e} = 0.2$ Gyr from the results of our hierarchical chemical evolution model and $M_{\rm max} = 10^{10}\msun$, which is the dark halo mass of the Large Magellanic Cloud. 
$M_{\rm min}$ is set at a value so that $b_{\rm m}(z, M_{\rm min}, v_{\rm rel}) = 0$ is satisfied. 
For $v_{\rm rel} = 10, 30, 100 \hbox{ km s}^{-1}$, the encounter rate is $0.015, 0.00057$, and $0.000012$, respectively. 
Massive halos are dominant contributors due to their large cross section.  
Even when $M_{\rm max} = 10^{11}\msun$, the encounter rate is $0.0042$ for $v_{\rm rel} = 30 \hbox{ km s}^{-1}$.

After the escaped stars fall onto the dark halo of the MW, they move around the MW with several hundred $\hbox{km s}^{-1}$. 
The kinematic energy of these stars is larger than the gravitational potential of sub-halos of the MW, 
and the cross section is zero.

\end{document}